%% file: main.tex
\documentclass[acmsmall]{acmart}
\startPage{1}
\setcopyright{acmcopyright}

\newcommand{\Secl}[1]{\label{sec#1}}
\newcommand{\Secr}[1]{Sec.~\ref{sec#1}}

\newcommand{\Figl}[1]{\label{fig#1}}
\newcommand{\Figr}[1]{Fig.~\ref{fig#1}}
\newcommand{\Tabl}[1]{\label{tab#1}}
\newcommand{\Tabr}[1]{Table~\ref{tab#1}}

\usepackage{xspace}

\newcommand{\tamarin}{{\sc \small Tamarin}\xspace} 

\newcommand{\revAdd}[1]{#1}
\newcommand{\revRem}[1]{}

\begin{document}

\title{It Takes a Village: Bridging the Gaps between Current and Formal Specifications for Protocols} 

\author[D. Basin]{David Basin}
\orcid{0000-0003-2952-939X}
\affiliation{
    \position{Professor}
    \institution{ETH Z\"{u}rich}
    \city{Zurich}
    \country{Switzerland}
}
\email{basin@inf.ethz.ch}

\author[N. Foster]{Nate Foster}
\orcid{0000-0002-6557-684X}
\affiliation{
    \position{Professor}
    \institution{Cornell University}
    \city{Ithaca}
    \state{NY}
    \country{USA}
}
\email{jnfoster@cs.cornell.edu}

\author[K. McMillan]{Kenneth L.~McMillan} 
\orcid{0000-0000-0000-0000}            
\affiliation{
        \position{Professor}
        \institution{University of Texas, Austin} 
        \city{Austin}
        \state{TX}
        \country{USA}}                   
\email{kenmcm@cs.utexas.edu}          

\author[K. S. Namjoshi]{Kedar S.~Namjoshi}
\orcid{0000-0002-6379-2442}            
\affiliation{
    \position{Distinguished Member of Technical Staff}
    \institution{Nokia Bell Labs}            
    \city{Murray Hill}
    \state{NJ}
    \country{USA}
}                   
\email{kedar.namjoshi@nokia-bell-labs.com}          

\author[C. Nita-Rotaru]{Cristina Nita-Rotaru}
\orcid{0000-0002-9649-6789}            
\affiliation{
    \position{Professor}
    \institution{Northeastern University}            
    \city{Boston}
    \state{MA}
    \country{USA}
}                   
\email{c.nitarotaru@northeastern.edu}
     
\author[J. M. Smith]{Jonathan M. Smith}
\orcid{0000-0000-0000-0000}            
\affiliation{
    \position{Professor}
    \institution{University of Pennsylvania}       
  \city{Philadelphia}
  \state{PA}
  \country{USA}
}                   
\email{jms@seas.upenn.edu}

\author[P. Zave]{Pamela Zave}
\orcid{0000-0002-6568-2052}
\affiliation{
   \position{Research Specialist}
   \institution{Princeton University}
   \city{Princeton}
   \state{NJ}
   \country{USA}
}
\email{pamela@pamelazave.com}

\author[L. D. Zuck]{Lenore D.~Zuck}
\orcid{0000-0003-3613-1208}             
\affiliation{
  \position{Research Professor}
  \institution{University of Illinois Chicago} 
  \city{Chicago}
  \state{IL}
  \country{USA}  
}
\email{zuck@uic.edu}          

\begin{abstract} \looseness=-1
Formal specifications have numerous benefits for both designers and users of network protocols. They provide clear, unambiguous representations, which are useful as documentation and for testing. They can help reveal disagreements about what a protocol ``is'' and identify areas where further work is needed to resolve ambiguities or internal inconsistencies. They also provide a foundation for formal reasoning, making it possible to establish important security and correctness guarantees on all inputs and in every environment. 

Despite these advantages, formal methods are not widely used to design, implement, and validate network protocols today. Instead, Internet protocols are usually described in informal documents, such as IETF Requests for Comments (RFCs) or IEEE standards. These documents primarily consist of lengthy prose descriptions, accompanied by pseudocode, header descriptions, state machine diagrams, and reference implementations which are used for interoperability testing. So, while RFCs and reference implementations were only intended to help guide the social process used by protocol designers, they have evolved into the closest things to formal specifications the Internet community has.

In this paper, we discuss the different roles that specifications play in the networking and formal methods communities. We then illustrate the potential benefits of specifying protocols formally, presenting highlights from several recent success stories. Finally, we identify key differences between how formal specifications are understood by the two communities and suggest possible strategies to bridge the gaps.

\end{abstract}


\acmJournal{JACM}
\acmVolume{68}
\acmNumber{8}
\acmYear{2025}
\acmMonth{8} 
\acmDOI{10.1145/3706572}

\maketitle

\input{intro}
\input{spec}

\input{examples}

\input{gap}

\input{discussion}

\input{conclusion}
\input{ack}
\bibliographystyle{acm}
\bibliography{main.bbl}

\end{document}

%% file: intro.tex
\section{Introduction}

Network protocols are fundamental to the functioning of modern society, as more and more services are provided online, including critical infrastructure. Historically, these services have mostly been implemented with wired networks that connect conventional computers. But with the emergence of wireless and cellular technologies, their reach has extended to IoT devices and sensors, robotic systems, and autonomous vehicles. The next generation of cellular technologies, often referred to as \emph{NextG}, will further expand the set of services that can be delivered over a network including real-time remote healthcare, industrial automation, precision agriculture, smart infrastructure, holographic communication, space-based communication, and more.

We are thus facing a future where a vast number of complex and critical services will emerge, and it is unrealistic to expect them to function correctly and be resilient to attacks in every scenario. It is often assumed that the well-understood Internet protocols that have been used are mostly correct and secure. This assumption is, however, false: vulnerabilities in ``old'' protocols, such as DNS, are frequently discovered and exploited. A review of the Common Vulnerabilities and Exposures (CVE) database reveals that many vulnerabilities stem from poorly specified and poorly understood specifications, leading to implementations that are prone to attacks.

Despite limited use of formal methods in the design of its protocols, the Internet mostly works, for some definition of ``works.'' With the transition to NextG technologies, significantly higher levels of assurance will be needed, especially for networks supporting safety-critical systems. The sheer number of emerging technologies and the potential risk to critical infrastructure and human lives demand that future protocols be designed, analyzed, and validated with care.

Formal methods offer methodologies for rigorous specification and verification of network protocols. Yet, current tools require substantial expertise to use, particularly for complex protocols. To date, most successful analysis efforts have been carried out using domain-specific tools and often by the developers of those tools. There is no single ``point of entry'' for understanding what tools are available, what inputs and expertise are required to use them, and what their capabilities and limitations are. Consequently, it is understandable why standards bodies like the IETF and IEEE do not routinely use formal methods to support their work.

This paper analyzes the gaps between the formal methods and networking communities. We first describe the differing views on specifications held in these communities. We then present examples where formal methods have been successfully used to improve real-world protocols, demonstrating how formal specifications and associated tools can provide a valuable design aid. Finally, we propose ideas for bridging the gaps in the future.

%% file: spec.tex
\section{On Specifications}%

The word ``specification'' means different things to different groups. The groups we focus on here are the networking community and the formal methods (FM) community.  

\subsection{Specifications in the Networking Community}

Today, networking protocols are defined by standards bodies such as the IETF, IEEE, 3GPP, etc. The ``specifications'' they develop usually consist of prose documents, annotated with pseudocode, header descriptions, state machines, message sequence charts, etc. Often the specifications are accompanied by a \emph{reference implementation}, which can be used to test for \emph{interoperability}. When flaws or inconsistencies are detected during testing, the specification and/or reference implementation are fixed. In practice, reference implementations are updated more often than specifications, so they often serve as de facto specifications of many protocols.

As an example, consider TCP, which is described in IETF RFC 793~\cite{rfc793_tcp}, originally published in 1981. A state machine diagram on page 22 of the RFC shows an ACK message being sent in response to a SYN message when in the SYN-SENT state. Meanwhile, in an examples on page 31, both SYN and ACK are shown as the response, rather that just ACK. 
A later update in RFC 1122 \cite{rfc1122}, published in 1989 states:
\begin{quote}
 the arrow from SYN-SENT to SYN-RCVD should be labeled with "snd SYN,ACK", to agree with [\ldots].   
\end{quote}
That is, for several years, TCP implementations used a specification that was consistent with an example rather than the protocol's state machine diagram given in the RFC. Over time, TCP has continued to be the subject of numerous RFCs---e.g., RFC 9293~\cite{rfc9293}, published in 2022, supposedly corrects all the inconsistencies and bugs discovered in over four decades of use of TCP.

Of course, RFCs and their accompanying social processes and reference implementations have been extremely successful---they are the key artifacts that have driven the development and evolution of Internet protocols over many decades. Yet, social processes and RFCs often leave ambiguities and gaps in protocol descriptions. These ambiguities and gaps can, in turn, give rise to serious security vulnerabilities, as illustrated by some of the examples in Section \ref{secexamples}. The application of formal reasoning can help to discover and resolve such issues. But formal reasoning requires precise definitions of protocols and requirements, both of which are typically unavailable.

IEEE standards, issued by the various working groups under the IEEE Standards Association, are also often used as ``specifications.'' Unlike RFCs, which  are at most a few hundred pages and coordinated by small groups, IEEE standards are often long and complex---e.g., the 802.11 standard for ``WiFi'' was developed by a group with over 300 members and runs to over 3500 pages. In many standards, aspects  are deliberately left vague or deferred to implementations. This is also the case with IETF RFCs, since looser standards generally provide more room for innovation in implementations.

Last but not least, are the ``specifications'' of the cellular protocols and architectures standardized by the 3$^{rd}$ Generation Partnership Project (3GPP), which is itself an association of seven telecommunications standards development organizations (ARIB, ATIS, CCSA, ETSI, TSDSI, TTA, TTC).
3GPP standards cover all aspects of cellular telecommunications technologies, including radio access, core network, and service capabilities. The 3GPP specifications also provide hooks for non-radio access to the core network, and for interworking with non-3GPP networks.
As with RFCs and IEEE standards, 3GPP specifications are often intentionally vague and usually do not include reference implementations. While open-source cellular platforms do exist, it is usually difficult to obtain high-level specifications from code.

\subsection{Specifications in the Formal Methods Community}
In the FM community, a specification is usually understood as an unambiguous description of a system and of its desired properties, both with a precise semantics. While the construction of formal specifications for protocols requires effort, this effort often pays off by providing benefits such as the following:

\begin{itemize} 
\item The process of formalization brings clarity and understanding to protocol design. It is common for gaps and inconsistencies to be revealed in the process of constructing a formal model.

\item A protocol works only under assumptions about its environment, which may include interactions with other protocols. It is common for the process of formalization to lead to a clear understanding of these interactions, and to bring hidden assumptions out into the open. 

\item A specification (even a partial one) can be used to automatically generate tests, which often reveal protocol errors that might not be detected by traditional interoperability testing of implementations. 

\item A specification may also be used to comprehensively analyze protocol properties, through formal proofs which show that desired protocol properties hold for \emph{all} scenarios. Such proofs are especially valuable when the properties go beyond functional correctness and capture security, performance, fault-tolerance, etc.

\end{itemize}

Many associate formal methods solely with verification. As C.A.R.~Hoare once said, ``A program that has not been specified cannot be incorrect, it can only be surprising.'' Unfortunately, the complete set of properties a protocol is expected to satisfy (also called ``intents'' or ``requirements''), including security and performance properties as well as environmental assumptions, is rarely stated explicitly, making complete formal verification effectively impossible. What is possible is to analyze certain aspects of implementations, although there are still significant challenges. For example, an implementation may deviate from the intent of the protocol designer, or the analysis may require having a specification for the environment. Even so, validating implementations using partial specifications and lightweight FM approaches that only model certain aspects of the environment is often useful.

In principle, formal specifications of networking protocols could be used to

prove certain requirements, including relating high-level specifications to low-level implementations. Formal specifications can point to bugs in the implementation as well as inconsistencies or disagreements on what the protocol ``is.'' When either a protocol or its implementations need to be changed, specifications can help guide refactoring and thus increase agility. In general, formal specifications can be a valuable aid for the social processes used to design protocols.

While formal specifications have many potential benefits, there is a long way to go to make them useful and usable by the networking community. Today, applying FM to real-world protocols requires considerable skill, as well as time and money. For networking researchers it requires specifying protocols in a completely new way, which will likely slow down certain parts of the design process. There are also social issues of network developers resisting change in common practices. In Section~\ref{sec:gap}, we will identify barriers that prevent FM from being widely used by the designers of network protocols. But first, we describe a few examples where FM has been applied successfully in networking.

%% file: examples.tex
\section{Examples of success stories}\Secl{examples}

 There have been several attempts by the formal methods community to capture RFCs and reference implementations as formal specifications. These attempts resulted in many successes, yet, they all required significant effort and have not produced a general methodology for ``reverse engineering'' intents from documentation or code. This is hardly surprising since, as stated above, RFCs are ambiguous, unclear, and sometimes contradictory. An individual implementation, of course, is not ambiguous. Yet, it reflects design decisions that, at best, restrict the intent of the RFC and at worst create vulnerabilities and errors.  
\begin{table}[htb]
\input{table}

\caption{Summary of verification tools mentioned in examples}\Tabl{tools}
\end{table}

 We present \revRem{five}\revAdd{six} examples here. The first two examples concern subtle security and privacy properties and were carried out by expert teams. The next three examples use light-weight FM for modeling and  testing. These methods can, with further development by the FM community, be adopted by practitioners. If adopted, it is highly likely that they would result in increased productivity for protocol designers. The last example discusses the use of FM in software-defined networks, including work that has connected formal specifications to real-world implementations. \Tabr{tools} summarizes the tools discussed in the examples.

\input{tls}
\input{5g}

\input{quic}

\input{chord}

\input{sip}

\input{sdn}

%% file: table.tex
\begin{tabular}{||l|p{6cm}|p{6cm}||}
\hline
Tool & Main Usage & Website \\
\hline\hline
Alloy & 
 A widely used language and analyzer for software modeling 
 & \url{https://alloytools.org}\\
CryptoVerif &  A protocol prover that provides for specifying the security assumptions on cryptographic primitives
& \url{https://opam.ocaml.org/packages/cryptoverif/}\\
F* & A general-purpose proof-oriented programming language that combines dependent types with proof automation based on SMT solving and tactic-based interactive theorem proving& \url{https://fstar-lang.org}\\
ProVerif & An automatic cryptographic protocol verifier in the Dolev-Yao model, based on representing protocols by Horn clauses & \url{https://bblanche.gitlabpages.inria.fr/proverif/}\\
IVy  & A language and a tool for specifying, modeling, implementing and verifying distributed protcols & \url{https://microsoft.github.io/ivy/language.html}\\
Rocq (Coq) & A proof-assistant for reasoning about software correctness & \url{https://coq.inria.fr}\\
SPIN & A tool for verifying temporal properties of software models & \url{https://spinroot.com/spin/whatispin.html}\\
\tamarin & A tool for  analyzing cryptographic protocols using  symbolic cryptography model  & \url{https://tamarin-prover.com}\\
Z3 & A widely used automated theorem prover and constraint solver for symbolic logic & \url{https://github.com/Z3Prover/z3}\\
\hline
\hline
\end{tabular}

%% file: tls.tex
\subsection{Transport Layer Security (TLS)}
One of the success stories in formally analyzing a widely-used protocol
is the formal analysis conducted on \revAdd{both versions 1.2 and 1.3 of} \revAdd{the Transport Layer Security (TLS) protocol}.\revRem{TLS, both versions 1.2 and 1.3.}
\revRem{The Transport Layer Security (TLS) protocol}\revAdd{TLS} is the main means
for securing communications on the Internet.
It was originally created  by Netscape in 1995 under the name SSL
and released as open source under the name
TLS 1.0. The protocol went through many changes with version 1.2 being used for
many years. Significant effort went into formally analyzing TLS, as discussed below.

In 2013, the authors of~\cite{tls12_fm_sp2013}
developed a verified reference implementation of TLS 1.2, which follows
the RFC specification \cite{rfc5246_tls12}. 
TLS is a very complex protocol with numerous configuration options
and deployment options, supporting many ciphers and backwards compatibility.
The authors had to make decisions in terms of what to support, given that
some configurations and some supported ciphers are known to be broken.
Their reference implementation included the two main components: the
authenticated stream encryption for the record layer and key
establishment for the handshake. 
The verification was done using F*.
The authors did not formally account for side channel attacks
based on timing and acknowledge that proving the absence 
of such attacks would require different tools.

The design of TLS 1.3 represented a big change and 28 drafts were created till
the design was finalized. One of the biggest changes was reducing the latency of the handshake protocol and supporting 0-RTT connection resumption,
as well as by default supporting perfect forward secrecy. A modular symbolic model of the TLS 1.3-Draft 21 release
candidate, with \tamarin verification of the security requirements claimed in the draft, are 
in~\cite{tls13_fm_ccs2017}

The focus of~\cite{tls13_fm_ccs2017} is the handshake protocol as it was essential for the overall
security of the communication. 
The analysis revealed an undesired behavior which had an impact
on the strong authentication guarantees in some implementations of the protocol. As this work was conducted during the design process of TLS 1.3, it had a strong influence on the final draft.

Around the same time, a verified symbolic and computational models of TLS 1.3-Draft 18 was presented in~\cite{tls13_draft18_sp2017}.
The models were specified using ProVerif and evaluated considering known and previously unpublished vulnerabilities that were considered during the design of TLS 1.3. The work in~\cite{tls13_draft18_sp2017} presents the first machine-checked cryptographic proof for a CryptoVerif model of TLS 1.3 Draft-18.

With respect to the TLS 1.3 finalized RFC, the most recent effort~\cite{tls13_record_sp2017} developed and verified a reference implementation of the TLS 1.3 Record Layer protocol and its cryptographic algorithms. The model was specified and verified using F*.

%% file: 5g.tex
\subsection{5G Authentication Key Agreement}
Another important standard is 5G, the current architecture and set of protocols for cellular networks. A
key component of the standard is the 5G Authenticated Key Agreement, or 5G-AKA for short, which describes how a \emph{user device} and a customer's \emph{home network} agree on a shared key.  
This protocol is critical for the security of communication on 5G networks as other keys are derived from this shared key, which are in turn used to protect user's data security, the authenticity of messages and
calls they receive, the connections they start, and even billing based on usage.

The work in~\cite{5g_fm_ccs2018} reports on a formal analysis of the 5G-AKA using \tamarin.
The protocol was formalized based on 3GPP's TS 33.501 document and the authors had additional access to 5G specialists.
The formalization was challenging given the 5G-AKA's complexity, both due to the specification's size and
its  usage in different contexts.  %
For example, when roaming, a user's device may connect to mobile networks (called {serving networks}) that are different from the service provider. The protocol then connects three parties, rather than just two, where only two of the parties, the user's device and home network, share initial secrets. 
Other complexities arise due to technological constraints and the need for backward compatibility.

The formal analysis started with an in-depth study of the relevant protocol documents, as well as discussions with some experts involved in the standardization. Afterwards,  an abstract version of the protocol was formalized in \tamarin.  The majority of the effort was the several person-months needed to understand the specification. In contrast, the time needed to subsequently formalize the resulting model was relatively short. Some additional person-months were needed for the verification of the protocol's security properties,
in particular writing proof strategies to help automate the construction of \tamarin proofs.

During verification, several flaws were found, which were
reported to the 3GPP.  The flaws were subsequently fixed 
in the standard, 
with one exception. The exception concerned the privacy of the
user's identity, which may be violated in the 5G-AKA protocol with a fairly simple
replay attack that exploits the 5G-AKA's resynchronization protocol with certain counters.
This problem cannot be easily fixed in the current design but a further iteration could solve this problem by shifting away from counter-based mechanisms. 

The most critical vulnerability discovered by this effort was a protocol error that allowed the attacker to induce confusion between users for the home network---i.e., the date or time
that are used and should be billed to customer $A$ would be incorrectly billed to
another customer $B$. The vulnerability was disclosed and fixed.
As a result, this verification work was
admitted to the ``GSMA Mobile Security Research Hall of  Fame'' as CVD \#0012, 2018.

%% file: quic.tex
\subsection{Quick UDP Internet Connection (QUIC)}

QUIC (Quick UDP Internet Connection) is a new Internet secure transport protocol. It has gone through over six years of IETF standardization, and its 35$^{th}$ version was declared as the standard in 2022 (RFC 9000). QUIC is intended as a replacement for the TLS/TCP stack and will be the basis of HTTP/3, the next official version of the hypertext transfer protocol.  It already carries a substantial amount of the traffic on the Internet, and is expected to carry even more in the future. 

QUIC's pivotal role rendered it a good candidate for the methodology of \emph{Network-centric Compositional Testing} (NCT), introduced by in~\cite{quic1,quic2}. The term ``network-centric'' refers to describing  a protocol's behavior as observed on the wire, rather than its abstract implementation. This approach is compositional, in the sense that any assumption made on the input of one process (or component) in the protocol can be treated as a requirement on the output of other processes (or components). Compositionality allows one to infer from the fact that implementations pass all tests that they will interoperate correctly when composed.  It also enables the discovery of cases where the specification is too strong (rejecting legal protocol behaviors) or too weak (peers misbehave when receiving an input the specification allows.)

The work reported in \cite{quic1,quic2} developed formal wire specifications for parts of QUIC, and tested implementations for compliance using these specifications. The testers take on one of the roles of the protocol. They generate packets that are legal according to the wire specification but may or may not be produced by any existing implementation. This generation can be done, for example, by using a modified SMT solver that randomly solves the constraints occurring in specifications, in such a way that every legal packet at a given point in the execution of a protocol has a non-zero probability of being generated.

This approach to testing has several advantages. It makes it possible to resolve ambiguities in informal standards documents using knowledge inherent in the implementations. At the same time, it verifies that the implementations comply with the formal specification as it is developed. It also exposes the implementations of a given protocol role to legal behaviors of the other role that may not be produced by existing implementations but may be produced by future implementations. Thus, it can detect interoperability issues before they occur in the wild.

Finally, the compositional nature of the NCT approach enables detecting cases when the specification is either too weak or too strong, allowing the developer to refine the specification accordingly. Consider a protocol with two roles $X$ and $Y$.  When testing an implementation of role $X$, the output of role $Y$ from the specification of role $Y$ is generated. If a response by the implementation of $X$ is not allowed by the specification of role $X$, then either the implementation is not compliant, or  the specification of $X$ is too restrictive.  On the other hand, if a response is not accepted by the implementation of $X$, then either the implementation of $X$ is not compliant, or the specification of $Y$ that allowed the response is too permissive.

The process of formally specifying QUIC uncovered numerous errors in the reference implementations, as well as some issues in the the draft standards themselves.  Examples of such issues include off-path denial of service attacks, one of which was a result of the standard as it appeared then in a draft RFC, and an information leak similar to the ``heartbleed'' vulnerability in OpenSSL. 

NCT was demonstrated on QUIC, but the method is quite general and has also been applied  to simpler protocols in classroom settings. The experiments used the IVy tool which in turn relies on the Z3 theorem prover. The size and complexity of QUIC packets presented challenges for efficient automated test generation. To overcome this problem, the specification must be carefully structured, a task that requires expertise on the part of the user~\cite{quic2}.

NCT has several features that may make it applicable to other complex protocols: (i) it allows testing of \emph{partial} specifications---i.e., one need not specify the whole protocol to enjoy the benefits of testing the parts that are captured in the formal model; (ii) the random nature of the generated tests allows testing of corner cases that interoperability testing rarely reveals; and (iii) the randomized tests, while not originally intended as such, turned out to be useful as adversarial inputs, thus allowing detection of security flaws. The experience of applying NCT to QUIC revealed that many security flaws can be detected using randomized testing, similar to how fuzzing has been effective for finding vulnerabilities in general-purpose software.

%% file: chord.tex
\subsection{Chord}
\label{sec:chord}

The Chord distributed hash table
was first presented in a
SIGCOMM paper which won
the 2011 Test-of-Time Award \cite{chord-sigcomm} for its authors.
An RFC based on Chord has also been developed
\cite{RFC-Chord}.
A Chord instance is a ring-shaped peer-to-peer network, and
the papers specify the protocol for maintaining the ring in
pseudocode and text.
Correctness of this protocol is a form of eventual consistency, meaning that the protocol
can eventually repair all defects due to failures, so that all live
members can continue to reach each other.
The SIGCOMM paper 
states, ``Three features that
distinguish Chord from many other peer-to-peer lookup protocols
are its simplicity, provable correctness, and provable performance''~\cite{chord-sigcomm}. 

However, despite these assurances, the work in~\cite{chord-ccr} shows that,
under the same assumptions made in the Chord papers,
the ring-maintenance protocol is not correct, and not one of
the seven properties claimed as invariants actually holds.
These results were obtained through automated checking
of a simple formal model in
the Alloy language.
Employees at Amazon Web Services
credit this work for motivating them to start using
formal methods~\cite{amazon}. 

Most of the seven claimed invariants are obviously useful and 
important, either for ring maintenance or for key lookup.
By fixing the flaws revealed by formal modeling and analysis, \cite{chordCorrect} shows it
is possible to specify the Chord protocol in a way that is both
correct and as efficient as the original.
The proof of correctness is particularly interesting, because 
although the specified protocol satisfies the claimed properties,
the properties do not resemble the true invariant upon which the
proof is based.

%% file: sip.tex
\subsection{{Session Initiation Protocol (SIP)}}
\label{sec:sip}

SIP is the dominant signaling protocol for IP-based audio,
video, and multimedia communication,
and has been standardized by the IETF.
The baseline SIP protocol is defined in a 268-page informal document~\cite{SIP-NRFC}.
Due to extensions, interoperation with other protocols, and
explanatory material, its description as of 2009
consisted of 142 documents totaling tens
of thousands of pages~\cite{hitch}.

With a complex protocol specification like this, it is not surprising that
people working with SIP spend many hours trying to answer basic questions
such as ``Can a protocol endpoint in state {\it S}
send a message of type {\it m}?''
In this context, \cite{under} provides a simple formal model
in Promela (the language of the model-checker Spin)
that proved to be an invaluable resource for software developers.
Model checking with Spin
verified that the model does not deadlock,
that it is complete in the sense of including every message that
can be received in every state, and that all parts of the model
are reachable.
The model provides quick and reliable answers to programming questions.
It has also been used in~\cite{b2bua} to generate automatically a large number of
test cases 

In addition to these fundamental uses, the model answered several
long-standing questions, as discussed below.

{\it Consistency.}
Regardless of how it is described, a protocol should be internally
consistent.  
SIP experts worry that SIP's many extensions have introduced 
inconsistencies, meaning that a new behavior in an
extension violates an assertion or constraint in an existing document. 
These experts are aware that inconsistencies could easily survive
the documentation and standardization process and 
in fact, their fears are well founded.
Even two of the earliest extensions were shown in~\cite{under} to violate
fundamental assumptions of the protocol.

{\it Complexity.}
A widely used protocol should not be unnecessarily complex.
Each capability should be generalized as much as is reasonable
and convenient, in preference to adding new capabilities that accomplish
similar and overlapping goals. The most important piece of SIP is the \emph{invite} 
transaction, a
three-way handshake allowing two endpoints to set up and negotiate the
parameters of a set of media channels between them.
Two early extensions serve
the same function as the invite transaction, in different but overlapping
circumstances.
The cost of these extensions is considerable.
They require five new message types, and a simplified model of just
one of them
requires 11 states and 15 state transitions \cite{under}.
Yet with the addition of a single Boolean flag to the messages of the
{\it invite} transaction, all of this additional complexity could be
avoided, and all media-control functions could be performed with
{\it invite} transactions alone.

{\it Race conditions.}
A race condition occurs when two messages cross in 
transit, either going the same direction (\Figr{races} left)
or different directions (\Figr{races} right).
Either kind of race can happen in SIP, especially when SIP messages
are sent via UDP instead of TCP
at the transport level (either is allowed).

\begin{figure}
\centering
\includegraphics[scale=1.0]{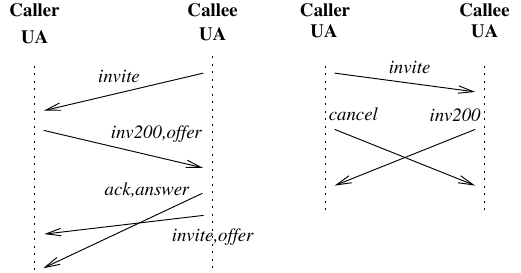}
\caption{Race conditions arise when messages going in the same or
opposite directions cross in transit.}
\Figl{races}
\end{figure}

Many SIP documents use message-sequence charts to show particular
common scenarios.
These charts are rendered in ASCII by means of IETF macros, and look
like \Figr{noraces}.
(Note that these charts represent message transmission as instantaneous,
so that race conditions are impossible.)
Not surprisingly, SIP race conditions are not well documented, and
their handling is incompletely standardized.

\begin{figure}
\centering
\includegraphics[scale=0.6]{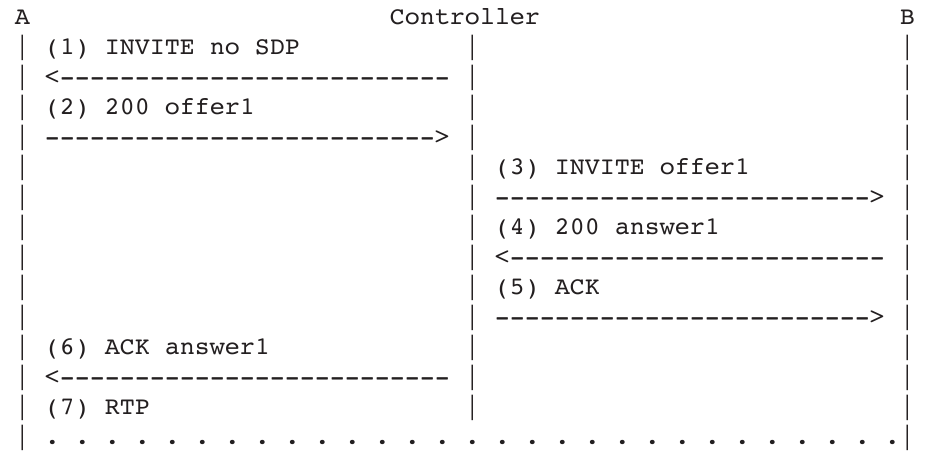}
\caption{Because the IETF standard for message-sequence charts
shows instantaneous message delivery, race conditions appear to be
impossible.}
\Figl{noraces}
\end{figure}
 
A SIP document \cite{races}, many years in the making, records 
seven possible
race conditions within the scope of the Promela model~\cite{under}.
The Promela model points to race conditions wherever they show that a
message can be received when it is not expected or desired.
By examining these places in the model, it is straightforward to find
the same seven race conditions and 42 others. Note that ater IETF documents, such as \cite{races},
show that ASCII art for race conditions has been developed.
The first specification of proxy (transparent) behavior for SIP
was also based on the formal model \cite{b2bua}.

%% file: sdn.tex
\subsection{Software-Defined Networks}

Software-defined networks and programmable networks more generally make it possible to formally specify and verify the \emph{low-level implementations} of network infrastructure. Given the number of interacting systems, the complexity of the software in network components, and the need for decentralized control, making network infrastructure programmable without also supplying ``power tools'' from FM is inherently risky. Thus, even the first instances of programmable networking (e.g., SwitchWare~\cite{switchware1998,securebootstrap1998} and Active Bridge~\cite{active_bridging}) were designed to eliminate large classes of errors by use of strongly-typed programming languages, such as Standard ML
and OCaml.

The emergence of SDN, which opened up interfaces for programming the network infrastructure, also led to significant interest in applying FM to network infrastructure. Building on early work that used  static analysis to analyze router configurations~\cite{xie}, tools such as Veriflow~\cite{veriflow}, Header Space Analysis~\cite{hsa}, NetKAT~\cite{netkat}, and Atomic Predicates~\cite{atomic-predicates} allow operators to specify and verify network-wide data plane properties, such as connectivity, loop freedom, way-pointing, etc. These early tools often worked with abstract models of network devices, but a number of recent tools have closed the gap with lower-level hardware models~\cite{p4v,p4testgen} and also explored applications of run-time verification~\cite{hydra}. A complementary line of research extended work on network verification to handle richer models of network functions, such as load balancers, firewalls, content caches, and so on~\cite{vigornat}. The distinguishing feature of these network functions is that they are written in general-purpose languages like C/C++ and typically rely on state---e.g., to track flows. Yet another line of work developed techniques for reasoning about conventional control-plane protocols, such as BGP and OSPF~\cite{batfish}.

%% file: gap.tex
\section{The Gap}
\label{sec:gap}
The community that develops Internet protocols aims towards correct and secure protocols. The current descriptions of the protocols do not allow for analysis, and thus for guarantees of correctness or security. FM is the most promising means for achieving these goals, but it requires formal specifications. Hence, there is a gap between the ways protocols are currently defined and the potential of FM to improve protocols throughout their life-cycles.

The purpose of this section is to point out specific 
challenges that must be addressed to fill the gap.
But first, we would like to note two issues that are often thought
of as obstacles to the use of FM, but are not:

\noindent\emph{(1) Incompleteness:}
Protocol specifications are usually incomplete, often for
good reasons: 
some details are best left to hardware and
software choices in implementations,
vendors want space to compete,
there are extensions to the original protocol, or there are simply
issues that were not well understood when the specification was written.
Because the essence of FM is abstraction, where
``abstraction'' is another way of saying ``incompleteness,'' formal
specification languages are actually far better than
implementation-oriented languages for this purpose.
A well-chosen specification language enables its users to state
essential facts while being perfectly clear about what it is
{\it not} saying.

\noindent\emph{(2) Scale:}
In the same vein,
scalability and agility are not barriers to obtaining the practical
benefits of formal specification in networking. Though informal
specification documents are often voluminous and complex, we have
found in practice that a small number of formal methods experts can
effectively capture the formalizations of complex protocols, such as QUIC,
and keep up with changes as a protocol evolves. Moreover, existing
tools have sufficient capacity to handle these specifications. There
are several reasons for this. First, the specification need not
include every detail, as noted above.
Second, and more importantly, these efforts do not
involve detailed proofs about code or hardware. Rather, they are
applications of {\it light-weight} FM. They use
relatively inexpensive techniques such as specification-based testing
to resolve ambiguities in informal documents, and reserve proof for
simple abstract models. This makes it possible to obtain the benefits
of formalization we have outlined at a moderate cost in human and
computational effort, albeit by sacrificing the ambitious goal of
certain correctness.
 
Nevertheless, significant challenges remain;
we have identified the following:

\noindent\emph{Semantic divide:}
    The FM community wishes for specifications that are rigorous and unambiguous. Such specifications serve as  thinking tools,  contracts between protocols and their environments (or even different parts of a protocol), as well as the basis for formal proofs, rigorous testing, and automated code extraction and synthesis. 
    
    The community specifying Internet protocols wish for consensus among the stakeholders, expediency, and flexibility for vendors who need to distinguish their products. This implies that the specifications of the protocols should be easy to read and understand.   Practitioners are more comfortable with natural language, despite its ambiguities, than with formal specifications.

    \noindent\emph{Finding the right tool:}
    Navigating among the multitude of tools available, and finding those that may be helpful, is challenging. The examples that we provided in \Secr{examples} highlight the problem: several different tools were used for modeling the protocols, and the analysis was conducted by researchers that were either the creators of the tool they used, or had significant experience with it. 
    The lack of commonality among tools impedes their developers as well. For examples of a  different approach, observe how agreement on the DIMACS SAT format spurred the competition to develop better SAT tools, and
    similarly for SMTlib. Similarly, Rocq (previously Coq) and Lean are becoming a de-facto standards, enabling the reuse of libraries of proofs across projects.

    \noindent\emph{Using the tools:}
    Even when practitioners find a suitable formal language and tool, and succeed in encoding their protocols
    as formal specifications, they find that the supporting tools require a considerable amount of expertise and training to use.
    Even with practice, the tools may be much more difficult to use than necessary.

    This is particularly unfortunate because good tools
    can do so much.
    They can make specifications more comprehensible with visualizations and alternative views.
    They allow analysis and verification, they can simulate behavior and generate test cases, along with other functions we have mentioned.
    Although it is always important to compare the formal semantics
    of a specification to the assumptions and expectations in
    its users' minds---the step called {\it validation}---experience
    shows that a variety of validation checks can be assisted
    with automated tools \cite{validation}.

    \noindent\emph{Limited support for quantitative properties:} Historically, much of the focus of the FM community has been on qualitative aspects of system behavior---e.g., functional correctness.
    However, important properties of network protocols are often quantitative or even statistical in nature, and there are very few tools to handle such properties. %
    Protocols are expected to have a behave ``well'' most of the time and be reactive to ever-changing environmental conditions.  There are very few formal frameworks, and, consequently, tools, that can deal with such complex properties.
    
   \noindent \emph{Limited support for security properties:} Security adds another dimension to the gap as  security properties are inherently difficult to formalize and check. Some require reasoning about ``hyperproperties'' (properties that involve several runs of the protocol), while others some require complex cryptographic analyses. All require reasoning about unknown adversaries including side-channels.
   
    \noindent\emph{Research Community Gap:}
    Experiments performed by Gerard Holzmann at Bell Labs in the 1980s (see \cite{Holzmann94}) demonstrated that systems developed using FM take less time to construct than systems built using a ``develop first, analyze later'' approach. While Holzmann's study only provides a single data point, recent work using FM in the design of cloud services and cryptographic hardware modules hints at the potential advantages of integrating FM into the design process. 
    Even so, it seems difficult to convince most developers that designing a system carefully and formally may actually save them time.  The prevailing wisdom appears to be that just writing  something up that (kind of) works and then debugging, testing, and reaching a social consensus that glosses over details and (kind of) captures what has been implemented seems faster. 

    The IETF (and 3GPP) develop their protocols through an iterative debugging process. This approach is unlikely to change. (The process also considers non-correctness factors such as messaging overhead, expected protocol convergence time, and support for multiple features.) It is the responsibility of the research community to create methodologies and tools that will convince designers and developers to use them while designing a protocol. Likewise, it is the role of the FM community to develop techniques that match the iterative development of networking protocols, and demonstrate, as in Holzmann's work, that formal specifications can considerably speed it up.

%% file: discussion.tex
\section{How to bridge the gap}

Given the large gap between the potential benefits of FM and the ability of protocol designers to use existing tools, we discuss several directions that can narrow this gap.

\noindent\emph{1. Engage industry~~} One way forward is to demonstrate the benefits of FM in industry, by finding bugs that really matter.
This was the case, for example, with work on the EMV payment protocol using \tamarin \cite{Conf:IEEE2021, Conf:Usenix2021}.  After finding serious flaws that allowed researchers to carry out high-value transactions without a PIN (the second authentication factor required for such transactions), the researchers began a collaboration with one of the EMV vendors that lead to a new EMV kernel (subprotocol), C-8, which was formally verified.

\noindent\emph{2. Develop tools that are easy to use~~} 
The FM community needs to work on building usable and useful tools. Whether a tool is ``usable'' or ``useful'' depends on the audience the tool is meant to serve. Hence, more collaborations between the two communities are needed, as well as with human-factors researchers.

\noindent\emph{3. Develop tools that address specific needs in the networking community~~} One of the problems of formal studies of networking protocols is the lack of a good model of the environment as well as precise requirements from the protocols that operate in these environments. For example, consider congestion control protocols, which have received significant attention in the networking community.  Without a good model of the environment it is virtually impossible to identify which of the many congestion control protocols should be used and when. Even for simple Additive-Increase/Multiplicative-Decrease (AIMD) protocols, such as TCP New Reno, a theoretical study identified pathological cases where the protocol gets stuck in its SlowStart state~\cite{ZuckWen24}. Given initial parameters, light-weight formal tools can help identify conditions on the environment that avoid this undesirable behavior.

\noindent\emph{4. Integrate design and verification~~}
The design process used for TLS 1.3 illustrated the benefits of performing formal analysis on intermediate drafts and not only on the final product.
However, incorporating FM into the development process requires a considerable investment of time, and currently requires experts to perform the analysis. This cost can be decreased by better tools that can be easily used by non-experts and should be amortized over the long term as fewer bugs and vulnerabilities will be present. Additional automated tools facilitating collaborations, detecting ambiguities and inconsistencies in textual specifications, and tracking changes are
needed (and some of them are already being used) to ensure the end-to-end integration of design and verification. Bringing together networking and formal methods researchers is essential in identifying the needs for such tools and ensure cooperation in changing the design process.

\noindent\emph{
5. Develop formal notations to express protocols and an ecosystem around them~~} There are at least two major research opportunities to address the semantic divide.
One is the development of a standard formal protocol notation (or notations). It is  typical for a  verification tool to define its own distinct input language, tailored to a class of protocols or to a type of verification method. As a result, models developed with different tools cannot be shared or interlinked, making it difficult to apply a combination of verification methods, as is often required for a complete protocol analysis.
The second opportunity is in exploiting the recent remarkable advances in natural language processing such as Large Language Models (LLMs) to automatically translate human-readable protocol specifications to a machine-analyzable formal notation~\cite{rfcnlp_sp2022}. A key challenge is to find ways to cope with the gaps, errors, and ambiguities that may be present in protocol specifications, while also coping with the artificial errors (e.g., hallucinations) that could be introduced during the translation. Success in addressing these challenges would have a revolutionary impact on improving the quality of protocol specifications. 

DARPA’s Open Programmable Secure 5G (OPS-5G) program developed techniques for automatically translating standards documents into precise executable models using natural language processing.
The highly structured nature of standards documents and their limited domain of discourse make them ideal targets of opportunity for natural language processing. Disagreements within a specification results in security holes as developers cope with differences in interpretation. Having an automated system for extracting an executable reference implementation from an informal specification provides a way to validate proprietary code developed by equipment vendors.

\noindent\emph{
6. Continue to integrate FM in computer science education~~}
FM has been increasingly recognized and integrated into computer science education. Many computer science programs have started integrating FM into their curricula, either as standalone courses or as components of existing courses such as software engineering, algorithms, or programming languages. 
Despite the progress, there are still challenges in teaching FM effectively. These challenges include the abstract nature of formal techniques, the steep learning curve for some concepts, and the need for specialized tools and expertise.

\noindent\emph{
7. Continue to increase awareness about the benefits of applying formal methods for network protocols~~}
We need to increase awareness of the benefits of existing formal analysis and tools in the networking community and beyond. For one, it seems that the networking community is not fully aware of the full benefits of FM. Perhaps methods that can highlight the potential benefits---not more examples to solve---can help to raise awareness. 
While summer schools were organized in the past, they focused on formal methods more generally. Perhaps a summer school or a Dagstuhl seminar with a focus on network protocols bringing together members of the two communities would help in this regard.

%% file: conclusion.tex
\section{Conclusion}
\label{sec:concl}

Formal methods offer methodologies and tools to rigorously specify protocols, verify parts of them, and even test their specifications and implementations. Numerous success stories demonstrate that these
benefits can be achieved in practice, reducing the costs associated with fixing bugs and vulnerabilities. However, several obstacles prevent their wider adoption and use by non-experts. In this paper we identified the gaps that prevent this adoption and described directions that can be taken to narrow the gap. 

%% file: ack.tex
\subsection*{Acknowledgements:}
 The authors would like to thank the anonymous reviewers for many helpful comments and suggestions for improvements. David Basin thanks the Werner Siemens-Stiftung for their generous support
of this project, under the Centre for Cyber Trust. Nate Foster was supported in part under NSF grant SHF-1918396, 
ONR contract N6833522C0411, DARPA 
contract HR001124C0429, and gifts 
from Juniper, Google, and the VMware University Research Fund. 
Kedar Namjoshi was supported in part by DARPA under contract HR001120C0159. Cristina Nita-Rotaru and Lenore Zuck thank the NSF in suggesting and encouraging this work, which was partially funded with grant CNS-2140207. The work of Lenore Zuck was also funded by the Discovery Partners Institute of the University of Illinois grant 634024 and by the NSF grant SHF-1918429. The views, opinions, and/or findings expressed are those of the authors and should not be interpreted as representing the official views or policies of any of the funding institutions and companies. 